# IMPEDANCE MEASUREMENT SETUP FOR HIGHER-ORDER MODE STUDIES IN NLC ACCELERATING STRUCTURES WITH THE WIRE METHOD *

N. Baboi [†], R. M. Jones, J. W. Wang, G. B. Bowden, V. A. Dolgashev, J. Lewandowski, S. G. Tantawi, P. B. Wilson, SLAC, Stanford, CA 94309

*Abstract*

Dipole modes are the main cause of transverse emittance dilution in the Japanese Linear Collider / Next Linear Collider (JLC/NLC). A diagnostic setup has been built in order to investigate them. The method is based on using a coaxial wire to excite and measure electromagnetic modes of accelerating structures. This method can offer a more efficient and less expensive procedure than the ASSET facility. Initial measurements have been made and are presented in this paper.

## 1 INTRODUCTION

Several types of accelerating structures are presently being investigated at SLAC for the JLC/NLC [1]. Two main issues are important for high energy physics colliders: achieving stable operation without electrical breakdown and minimizing long-range wakefields. In particular dipole wakefields are the main cause of emittance increase and beam break-up (BBU) in high energy accelerators. They have to be well understood in order to be able to damp them properly and thus obtain a high luminosity at the collision point.

For the experimental study of wakefields at SLAC, the ASSET facility has been used in the past [2]. There, a witness bunch is deflected by the fields excited by a leading bunch, sampling in this way the wakefield. Although it has proved to be a very reliable method, in good agreement with the circuit model theory [3], a stand-alone setup to measure dipole wakefields that would not require the use of a beam is desirable. The wire method fulfills this requirement.

This method consists of sending a current pulse through a wire which runs through an accelerating structure under test [4,5,6]. One can measure the distortion of the current pulse in time domain, in order to obtain information directly on the wakefields, or the scattering parameters in frequency domain. In order to measure dipole wakes, the wire is displaced from the axis. Alternatively two wires placed symmetrically about the axis can be used [7].

Two measurements are needed: one with the device under test (DUT) and one with a reference pipe. From these two measurements one extracts information regarding the impedance of the modes of the DUT [5,6]. From the integration of the impedance one can then calculate the modal loss factors. At SLAC a wire setup for measurements in the frequency domain has been built and is under test [8]. Initial measurements with a single-cell cavity are presented in this paper.

## 2 MEASUREMENT SETUP

The measurement setup is shown in Fig. 1. Two matching sections shown schematically in the figure are placed at either end of the DUT in order to minimize losses of the signal from the network analyzer. Fig. 2 shows in detail a matching section. It consists of a broadband waveguide-coaxial matching section (W) and a removable coaxial-wire adapting part (C). The adapter pipe is flexible enough to enable the movement of the wire in the transverse plane. This is achieved by moving the whole section head while the other end of the pipe is held fixed.

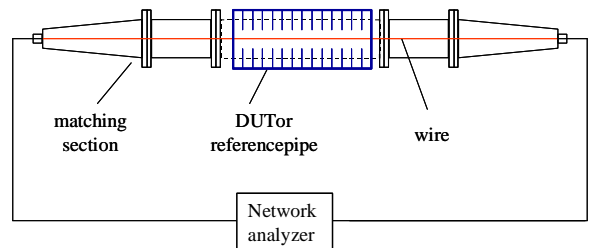

Figure 1. Wire measurement setup

The matching sections are broadband, the design S11 being below –30 dB from 11 to 18 GHz. It is anticipated that our experimental setup will be able to measure the first two dipole bands of the standing wave structures and the first band for the traveling wave structures [9].

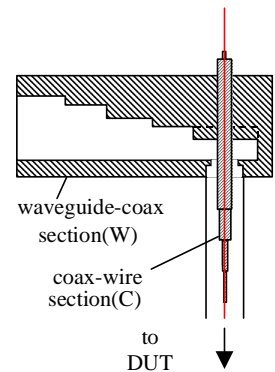

Figure 2. Matching section: is made of two detachable parts: waveguide-coax (W) and coax-wire (C)

___________________________

*Supported by the US Department of Energy, grant number DE-AC03-76F00515

[†] On leave from NILPRP, P.O. Box MG-36, 76900 Bucharest, Romania

The matching sections have been first investigated by connecting them directly with one another. A rod is used rather than the C matching sections, in order to analyze the W sections separately. From measurements in frequency domain using a network analyzer it could be seen that one of the waveguide-coaxial matching sections did not meet the design specifications. W1 presented a reflection coefficient of –18dB, while W2 meets the design goals, having a reflection of about –40dB over most of the frequency range. When the rod was replaced with the wire and the C-matching sections, the reflection from each adapter increased to about –20dB. HFSS simulations with an increased radius of the last step of the coax-wire adapter led to a reflection coefficient degraded by –20dB. This increased radius occurred due to a fabrication error.

Utilizing a Fast Fourier Transform the time domain signal was obtained. Fig. 3 illustrates the reflection measured from matching section 1. Reflections above 0.1 can be observed from each adapter. A small reflection is seen from the flange connecting the two adapters (at 2ns). A coupling between the two reflections induces a frequency modulation of the signal. These matching sections were used for initial measurements, while more precise ones are in the process of being fabricated.

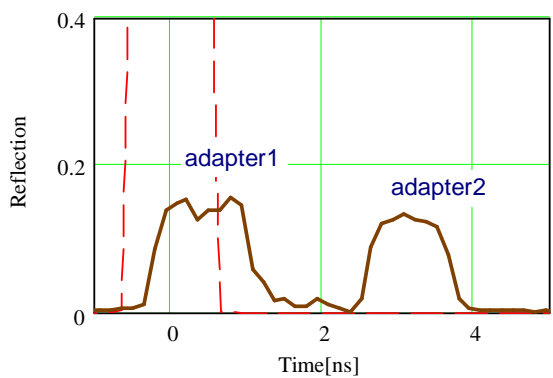

Figure 3. Time domain measurement of the matching sections. The dashed curve represents the input pulse.

## 3 INVESTIGATION OF SINGLE CELL

### Measurements

A copper single cell has been mounted in the measurement setup. The cell has a radius of 10.7mm and a length of 9.8mm. The pipe itself is made of stainless steel and has a radius of 4.7mm. The S parameters have been measured over the frequency range between 11 and 18 GHz with the DUT as well as with a reference pipe having the same length as the cavity, using a calibrated network analyzer. The wire has been approximately centered in the system, by centering the ends of the matching sections in the fixture system. A more precise transverse alignment of the wire has been then achieved by minimizing the amplitude of the transmission of a dipole mode of the cavity. The wire has been then displaced in 250 μm steps, to up to 2mm offset from the axis. Fig. 4 presents several measurements at increasing wire offsets. Both the measurement with the DUT (solid lines) and the one with a reference pipe (dashed lines) are shown. The dot-dashed lines will be explained in the next section. The maximum transmission is below unity, due to the attenuation in the coaxial line with copper central conductor and stainless steel walls and in the adapters. By comparison of the two curves in each plot it can be seen that most of the losses occur in the matching sections and beam pipe, the ones in the cavity being negligible in comparison.

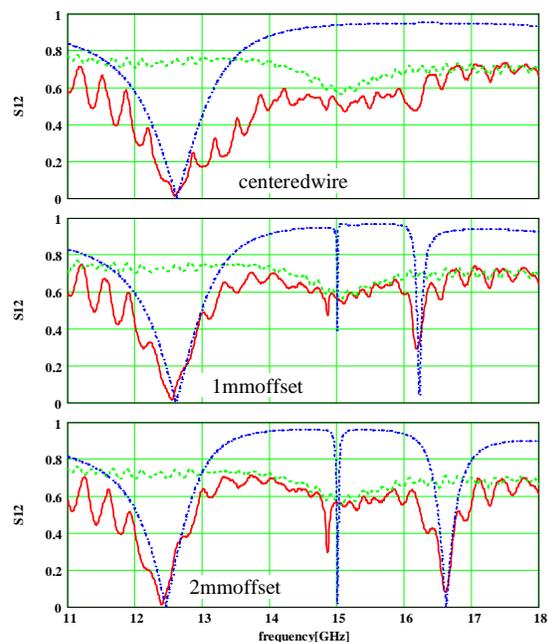

Figure 4 Transmission parameter $S_{12}$ for various wire offsets. The curves represent the measurement with reference pipe (solid line), measurement with DUT (dashed line) and HFSS simulation of cavity with lossy wire (copper) and walls (stainless steel) (dot-dashed line)

A frequency modulation is observed in all measurements. This is due to the mismatch in the adapters shown in the previous chapter. A broad peak representing a monopole resonance in the cavity can be observed when the wire is centered in the device. When the wire is displaced by 1mm, two higher quality resonances appear at about 14.8 and 16.6 GHz. They represent the first dipole modes of the cavity, TE-like and TM-like, respectively. The frequency of the first dipole mode stays constant while moving the wire, while the other shifts to higher values. On approaching the cell axis, the frequency approaches the frequency of the mode in the unperturbed cavity.

### Simulations

The S parameters have been calculated for a single cell attached to tubes at either end utilizing HFSS (High Frequency Structure Simulator), in order to understand the measurements. Both the wire and the cavity walls are assumed to be lossy (copper wire and stainless steel walls). S12 obtained from simulation is compared with the data in Fig. 4. The amplitude of the signal is much closer to unity since losses in the matching sections are not taken into account. Despite the modulation in the measured data, there is a clear correspondence between measurement and simulation. A discrepancy is observed in the frequency of the first dipole mode.

Fig. 5 presents the frequency of the modes as a function of the wire offset. The crosses represent the measurement points and the circles the simulation results. The horizontal dot-dashed lines mark the frequencies of the dipole modes in the unperturbed cavity. Good agreement between simulation and measurement up to 50 MHz is seen in the monopole and the second dipole mode. A shift of approximately 150 MHz is observed in the first dipole mode. A tilt in the wire or a coupling in resonances may explain this. Further studies are in progress to understand this discrepancy.

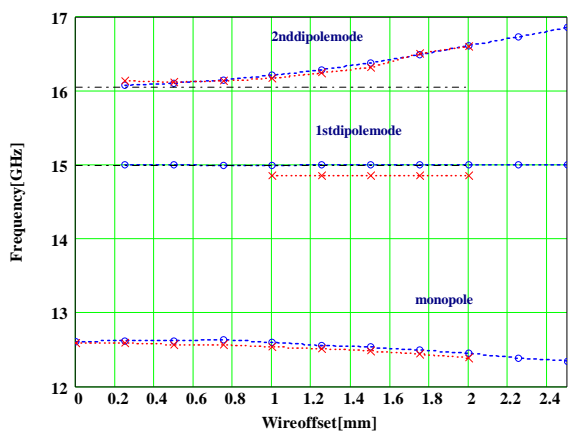

Figure 5 Synchronous mode frequencies as a function of the wire offset with respect to the axis: measurements (crosses) and simulation (circles). The horizontal dot-dashed lines represent the frequencies of the dipole modes of the cavity without perturbing wire.

The electric field of the dipole modes is shown in Fig. 6. The TE-like mode almost automatically fulfills the boundary condition on the wire surface. The wire only modifies the field distribution locally. The perturbation is therefore small and the frequency is not changed. In the case of the TM-like mode, the wire forces the field lines to change, inducing a larger perturbation. The frequency shift increases when the wire is moved into regions with higher field amplitude. We note that in the case of the monopole mode (TM-like) the maximum frequency change is when the wire is in the middle of the cavity, where the field is also maximum (the unperturbed frequency is 11.2 GHz).

The property of the frequency variation with the wire offset for only the second dipole mode, while the one of the first one remains constant is characteristic for single cells. For multi-cell cavities, as is the case of most accelerating RF structures, the modes are rarely TE-like or TM-like, but rather hybrid modes. Therefore the frequency will shift in both modes [8]. By extrapolating the frequency curve at the axis, one can obtain the mode frequency.

In order to characterize the wakefields it is important to have knowledge of the modal loss factors. The loss factors of the modes were calculated with HFSS for the unperturbed cavity as well as in the presence of the wire. When the wire approaches the cavity axis the loss factor equals the value corresponding to the unperturbed structure. The loss factors are currently under investigation and will be reported in a later publication.

## 4 SUMMARY

A setup to study dipole modes based on a wire method has been recently been built and is currently under test at SLAC. A test cell has been mounted in the experimental setup and initial measurements have been made. The measurement is strongly influenced by the mismatches in the matching sections. Nevertheless, agreement with HFSS simulations has been observed. The frequency of the first dipole mode is unperturbed by the wire due to the TE-like character, while the TM-like dipole mode gives rise to a shift in frequency with the wire offset. This shift becomes zero as the wire is brought to the cavity center. Further studies are in progress.

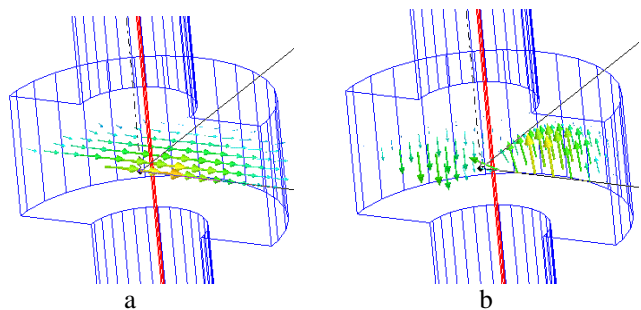

Figure 6 Electric field of the two dipole modes: TE-like (a) and TM-like (b)